\documentstyle[amssymb]{article}
\begin{document}
\noindent

{\large \bf Unified  $ (p,q; \alpha,\gamma, l)$-deformation of
oscillator algebra and  two-dimensional conformal field theory
} \vskip 10 pt

{ I. M. Burban}

{\it Bogolyubov Institute for Theoretical Physics, Nat. Acad. of
Sci. of Ukraine}

{(14 b, Metrolohichna Str., 03680 Kyiv, Ukraine)}

\vskip  40 pt
\noindent
{\bf Abstract}

The unified $ (p,q; \alpha,\gamma, l)$-deformation of a number of well-known deformed oscillator algebras is introduced.The deformation is constructed by imputing new free parameters into the structure functions and by generalizing the defining relations of these algebras. The generalized Jordan-Schwinger and Holstein-Primakoff realizations of the $U_{pq}^{\alpha \gamma l}(su(2))$ algebra by the generalized $ (p,q; \alpha,\gamma, l)$-deformed operators are found.

The generalized $ (p,q; \alpha,\gamma, l)$-deformation of the two-dimensional conformal field theory  is established.
 By introducing the $ (p,q; \alpha,\gamma, l)$-operator product expansion (OPE) between the energy momentum tensor
and primary  fields, we obtain the $ (p,q; \alpha,\gamma, l)$-deformed centerless Virasoro algebra.
The  two-point correlation function of the primary generalized $ (p,q; \alpha,\gamma, l)$-deformed  fields is
 calculated.

\medskip
\noindent
PACS: 02.10.Tr, 02.30.Gp, 03.30.Tb, 03.65.Dp

\medskip
\noindent
{\it Keywords:} Generalized deformed oscillator, structure function, generalization Jordan-Schwinger and Holstein-Primakoff transformations, deformed two-dimensional conformal field theory.
\medskip

Fax : 38(044) 526-59-98
\medskip

E-mail : burban@bitp.kiev.ua

\vskip 35 pt
\noindent{\bf 1. Introduction}
\medskip

\noindent

The quantum deformations of the universal enveloping  algebra $U_q(\cal L)$ of a simple  Lie algebra ${
\cal L}$ emerged in the study of solutions of the quantum Yang-Baxter equations \cite{F}.
Drinfeld extended this structure to general class associative algebras with the Hopf algebra structure,
which is neither commutative nor co-commutative. The non-co-commutativity is achieved by the introducing of
a formal parameter $q.$
An example of such structures is the deformation of the universal enveloping algebra $U_{q}\Bigl(su(2)\Bigr)$  \cite{S, KR}.
The important tool in the study of these structures is their realization by the creation and annihilation operators (the Jordan-Schwinger and the
Holstein-Primakoff constructions). In order to generalize the Jordan-Schwinger  construction to quantum algebras, Biedenharn \cite{B}  and  Macfarlane \cite{Mc} introduced, independently, {\it the q-deformed creation and the annihilation operators.}

Long before  another $q$-deformation of the canonical commutation relations has been used by Arik and Coon \cite{AC} for the operator description of the
generalized Veneziano amplitude obtained by replacement of the ${\Gamma}$-function by the
$\Gamma_q $-function.

Naturally, it would be desirable to generalize a particular mathematical structure as much as possible. In particular, it concerns the generalization of the one-parameter deformations of the Lie algebras, initiated in the works \cite{M, R, Br, SWZ}.

The $(p,q)$-deformed of the $U_{p,q}\Bigl(su(2)\Bigr)$ algebra and its realization
by the operators of the  $(p,q)$-deformed oscillator algebra was studied in refs.\cite {Ch-J, J}. A construction two-parameter oscillator algebra also has been introduced in \cite {ADT} under the name "Fibonacci" oscillator algebra.

 The sequently investigations have shown that for algebraic structure of the $(p,q)$-deformed quantum $U_{p,q}\Bigl(su(2)\Bigr)$-algebra the additional parameter is artificial and could be removed from it.
 On the level of oscillator algebra, the study of  multi-parameter deformations of the oscillator algebra was continued in the works  \cite{CU, BDY, MARC, MCD, Burba}.

All these deformations are  three-parameter generalizations of the one-parametric deformation of the Biedenharn-Macfarlane algebra. They
 yield the Jordan-Schwinger realizations of the generalized deformed Lie algebras $U_{p,q}\Bigl(su(2)\Bigr).$

 The multi-parametric generalization of the two-parametric deformed oscillator algebra was considered \cite{Burba,GH}. Although algebraic structure of any multi-parameter deformed quantum algebra may be mapped one-to-one onto the standard one-parameter deformed algebra \cite {CZ,P}, its co-algebraic structure and physical results in both cases is not the same.

 Many versions of the deformed oscillator algebras have been considered in the literature.  Most of them can be embedded within the common mathematical  framework of the {\it generalized deformed oscillator algebras.}

 A generalized oscillator algebra is an associative algebra  generated  by the generators $\{ 1, a , a^+, N \},$ where  $a$ and $a^+$ are hermitian conjugate $N$ self-adjoint operators, and the defining relations
\begin{equation}\label{burban: gener}
aa^+= f(N+1),\quad a^+a  = f(N),\quad [N, a] = - a,\quad [N, a^+] = a^+.
\end{equation}
 A positive analytic function $f(x) = [x]$ with $ f(0) = 0$ is  called {\it a structure function} \cite {D}.
This function defines a deformation scheme, and along with defining relations it entirely defines the deformed oscillator algebra.
The most well-known structure  function is the one of  the two-parameter deformed oscillator algebra
\begin{equation} \label{burban: func}  f(x)  = [x]_{pq} = \frac{p^{-x} - q^{x}}{ p^{-1} - q},
\end{equation}
$p,q \in {\mathbb R}$ \cite{Ch-J}.
The structure function (\ref{burban: func}) and defining relations
\begin{equation}\label{burban: firsta}
aa^+- q^{-1} a^+a = p^{N},\quad [N, a] = - a, \quad [N, a^+]=a^+,
\end{equation}
 or
\begin{equation} \label{burban: firstb} aa^+ - q a^+a = p^{- N },\quad [N, a] = - a,\quad [N, a^+] = a^+.
\end{equation}
define the $(p,q)$-deformed oscillator algebra \cite{Ch-J, J, ADT}.
 This algebra describes the Arik-Coon  $(p = 1, q)$ \cite{AC}, the Biedenharn-Macfarlane $(p = q^{\alpha},q^{\gamma}),$\cite{B, Mc}, the Kwek, Oh $(p = q^{\alpha},q^{\gamma})$ \cite{KWO} deformations of the oscillator algebras in unified framework.

The other version of the deformed oscillator algebra, different from (\ref{burban: firsta}) (or (\ref{burban: firstb})),
with the structure function
\begin{equation}\label{burban: func1}
 f(x)  = [x]_q = \frac{q^{-x} - q^{x}}{ q^{-1} - q}
\end{equation}
and the defining relations
\begin{equation}\label{burban: gener2}
[a, a^+] = [N+1]_q - [N] _q, \quad  [N, a] = - a, \quad [N, a^+] = a^+
\end{equation} is the oscillator algebra, which in general is distinguished from the previous ones.
The oscillator algebra (\ref {burban: firsta}) (or (\ref{burban: firstb})) and its generalizations have not of the Hopf algebra structure whereas the (HY) algebra is endowed
with this structure (HY) \cite {Y}.

The number of the free parameters of the structure function of the  deformed of oscillator algebra characterize a deformed scheme.
One of the methods to obtain a multi-parameter deformed algebra is to add to the structure function several the new parameters.
 Naturally, the increasing of the number of the deformation parameters makes the application of the deformed oscillator algebras more
flexible in the practical applications.
In the framework of one-parameter deformation this scheme combined with scheme of construction of the generalized deformed oscillator algebras
\cite{D} has given possibility to unify in unified framework \cite{CU, BDY, KWO, MARC, Burba} the  well-known deformations of the oscillator algebra \cite {AC, B, Mc}.

These generalize deformed oscillator algebras have found the application in the concrete physical models \cite{BD,MCD}.
The  authors \cite{BD}  have shown that the algebra of the observable system of two identical vortices in superfluid thin film is described by the generalized deformed algebra.
The authors \cite {MCD} considered this generalized deformed oscillator  and  have shown  that Kerr medium transforms an ordinary linear harmonic oscillator into this generalized
oscillator for specific values of the deformation parameters $\alpha,\gamma.$

The different deformation schemes of the oscillator algebras have found their parallel reflection in the constructions of concrete deformed models of quantum  physical systems. For example, the Jaynes-Cummings models in quantum optics, in parallel to deformed oscillator algebras, generated  the Biedenharn-Macfarlane's \cite{CEK},the f-oscillator's \cite{SSR}, the $(p,q)$-deformed versions \cite{Ch-J}  of this model.
The same takes place for two-dimensional conformal fields theories induced by  the Arik-Coon \cite{Ber},  the Biedenharn-Macfarlane \cite{Mat}, the $(p,q)$ \cite{C-J}, the Oh-Sigh \cite{OhS} deformed  versions of the oscillator algebras. In framework of the $ (p,q; \alpha,\gamma, l)$-deformation these theories allow unification.

The $q$-deformed oscillator algebras were used in the construction of the  three-dimensional non-commutative spaces with broken Lorentz invariance \cite{DFG, DFGC}.
The $ (p,q; \alpha,\gamma, l)$-deformed oscillator algebra may be useful in this area.

In this paper we  propose the unified deformation of the oscillator algebra which envelop  as particular cases the well known deformations \cite {AC, B, Mc,Ch-J, CU, BDY}.
The structure function ("generalized $(p,q; \alpha,\gamma, l)$-number") of this deformation
\begin{equation}\label{burban: func1}
f(x)  = [x]_{pq}^{\alpha\gamma l} = \frac{p^{-\alpha x} - q^{\gamma x}}{p^{-l/\gamma} - q^{l/\alpha}},
\end{equation} where  $ \alpha,\gamma, l \in {\mathbb R},$ contains comparatively with (\ref {burban: func}) the additional deformation parameters.

Traditionally at the fixed structure function, the defining relation of the generalized deformed oscillator algebras are represented in the form (\ref{burban: gener}).
We generalize the defining relations (\ref{burban: firsta}) (or (\ref {burban: firstb})) and (\ref{burban: gener2}) to obtain the various generalized deformed versions of the oscillators at the fixed structure function (\ref{burban: func1}). We consider only generalized Daskaloyannis (GD),  generalized Chakrabarti-Jagannathan (GCh-J), and  generalized Hong Yang (GHY) versions of these  algebras.
The properties of these oscillator algebras in general are distinguished from each others. We study the relationship between them.
The realization of the $U_q\Bigl(su(2)\Bigr)$  algebra  \cite{S, KR} by the $q$-deformed and $(p,q)$-deformed creation and annihilation operators independently
has been done by different authors \cite{B, Mc, Ch-J, J, KWO}.

Generalizing the Jordan-Schwinger and the Holstein-Primakoff  transformations we obtain the realizations $U_{pq}^{\alpha\gamma l}\Bigl(su(2)\Bigr)$ algebra in terms of operators of the (GCh-J) and (GHY) deformed oscillator algebras.
For convenience in the physical applications we included in formulas representations the non-zero Casimir invariants of the generalized deformed (GCh-J) and (GHY) oscillator algebras.

The above deformations of the oscillator algebra have found applications in the construction of the deformed two-dimensional integrable models, and the two-dimensional conformal field theory. Some of these deformations of the two-dimensional conformal  field theories have been studied in  the works \cite{Mat, Ber, OhS, Chu}. The problem of the unification of these deformed theories arose also in this case.

We have considered the $ (p,q; \alpha,\gamma, l)$-deformation of the two-dimensional conformal field theory.
Note from the beginning, in sofa as the satisfactory central extension of the quantum deformed complex Witt algebra does not exists we restrict ourselves  by the consideration of the centerless  $ (p,q; \alpha,\gamma, l)$-deformed Virasoro algebra and by the investigation of the two-dimensional conformal field theory basad on it.
In the following the $ (p,q; \alpha,\gamma, l)$-deformed conformal field theory means the conformal field theory based on centerless $ (p,q; \alpha,\gamma, l)$-deformed Virasoro algebra.

The conformal transformations of primary fields are
generalized to $(p,q; \alpha,\gamma, l)$-deformed conformal transformations. We have found the pole structure of the $ (p,q; \alpha,\gamma, l)$-deformed operator product expansion (OPE)  of the holomorphic component of the energy-momentum tensor with primary fields. From this we have obtained centerless $ (p,q; \alpha,\gamma, l)$-deformed Virasoro algebra.
We have found the two-point correlation function of the  $ (p,q; \alpha,\gamma, l)$-deformed two-dimensional conformal field theory.

\medskip

\noindent {\bf 2. Some aspects of generalized $ (p,q; \alpha,\gamma, l)$-deformed oscillator algebras}
\medskip

 At the fixing structure function (\ref{burban: func1}) we will consider the basic versions of the generalized deformed oscillator algebras and the relations between them.
\noindent

{\it The generalized $ (p,q; \alpha,\gamma, l) $-deformed Daskaloyannis version of the oscillator algebra.} In this case the defining relations
(\ref{burban: gener}) are written
\begin{equation}\label{burban: qgener}
aa^+ = \frac{p^{-\alpha N - l/\gamma} -  q^{\gamma N + l/\alpha}}{p^{-l/\gamma} - q^{l/\alpha}},
a^+a  = \frac{p^{-\alpha N } - q^{\gamma N}}{p^{-l/\gamma} - q^{l/\alpha}},
 [N, a] = - \frac{l}{\alpha\gamma} a, [N, a^+]=\frac{l}{\alpha\gamma} a^+.
\end{equation}
\noindent {\it The generalized  $(p,q; \alpha,\gamma, l)$-deformed Chakrabarti-Jagannathan version of oscillator algebra.}  In \cite {Burba} we have been generalized two-parameter
deformed oscillator algebra \cite {Ch-J} by introducing additional parameters $\alpha, \gamma, l$ in the structure function. This is a generalization of the well-known deformations \cite {AC, B, Mc, Ch-J, CU, BDY, KWO} of these algebras.
The  generators $\{I, a, a^+, N\}$, structure function (\ref {burban: func1}) and the defining relations
\begin{equation}\label{burban: qfirsta}
 aa^+- p^{-l/\gamma} a^+a = q^{\gamma N},\quad
 [N, a] = - \frac{l}{\alpha\gamma}a, \quad [N, a^+] =\frac{l}{\alpha\gamma} a^+,
 \end{equation}
 or
 \begin {equation} \label{burban: qfirstb} aa^+ - q^{l/\alpha} a^+a = p^{-\alpha N},\quad
 [N, a] = - \frac{l}{\alpha\gamma}a, \quad [N, a^+] = \frac{l}{\alpha\gamma} a^+,
  \end{equation}
  define the generalized $(p,q; \alpha,\gamma, l)$-deformed  Chakrabarti-Jagannathan version of the oscillator algebra.

 It is easy to see that this algebra is represented in the Hilbert space ${\mathbb H}$ with the basis $\{ |n\rangle \}_{n=0}^{\infty}$ by
 \begin{equation} a|0\rangle= 0,\quad |n\rangle =([n]_{pq}^{\alpha\gamma l}!)^{-1/2}(a^+)^n|0\rangle,
 \end{equation}
where $[n]_{pq}^{\alpha\gamma l}$ is given by (\ref{burban: func1}) and $[n]_{pq}^{\alpha\gamma l}! =[n]_{pq}^{\alpha\gamma l}[n-1]_{pq}^{\alpha\gamma l}[n-2]_{pq}^{\alpha\gamma l}\ldots[1]_{pq}^{\alpha\gamma l}.$
 Then
$$
N |n\rangle = n |n\rangle,
$$
 \begin{equation}
a|n\rangle = \Bigl(\frac{p^{-\alpha n} - q^{\gamma n}}{p^{-l/\gamma} - q^{l/\alpha}}\Bigr)^{1/2} |n - \frac{l}{\alpha\gamma}\rangle,\quad
a^+|n\rangle =
\Bigl(\frac{p^{-\alpha n -l/\gamma} - q^{\gamma n + l/\alpha}}{p^{-l/\gamma} - q^{l/\alpha}}\Bigr)^{1/2} |n + \frac{l}{\alpha\gamma}\rangle,
\end{equation}
where $ p,q, \alpha,\gamma \in{\mathbb R}$  and $\frac{l}{\alpha\gamma}\in {\mathbb Z},$
 of this deformed algebra defined by the representation of this algebra in the Fock representation. The other version of this generalized deformed oscillator algebra is

\noindent {\it The generalized  $(p,q; \alpha,\gamma, l) $-deformed
Hong Yan version of oscillator  algebra.} The  generators $\{I, a,
a^+,N\}$, structure function (\ref {burban: func1}) and the defining
relations
$$
aa^+- (p^{-\alpha} q^{\gamma})^{\frac{l}{2\alpha\gamma}}   a^+a =
\frac{p^{-\alpha N -\frac{l}{2\gamma}}+q^{\gamma N
+\frac{l}{2\alpha}}}
{p^{-\frac{l}{2\gamma}}+q^{\frac{l}{2\alpha}}},$$
\begin{equation}\label{burban: qforth}
N a -a N = - \frac{l}{\alpha\gamma} a,\quad  N a^+- a^+ N =\frac{l}{\alpha\gamma} a^+
\end{equation}
define the generalized $(p,q; \alpha,\gamma, l) $-deformed Hong Yan (GHY) version
of oscillator algebra.
The first relation  in (\ref{burban: qforth}) can be rewritten in the form
\begin{equation}\label{burban: diff}
aa^+- (p^{-\alpha} q^{\gamma})^{\frac{l}{\alpha\gamma}}
  a^+a = [N+\frac{l}{\alpha\gamma}]_{pq}^{\alpha\gamma l} - (p^{-\alpha}q^{\gamma})^{\frac{l}{\alpha \gamma}}[N]_{pq}^{\alpha\gamma l}.
\end{equation}
The relations (\ref{burban: qgener}), (\ref{burban: qfirsta}), (\ref{burban: qfirstb}), (\ref{burban: qforth})
define the three versions of the generalized $(p,q; \alpha,\gamma, l)$-deformed oscillator algebra.
It is easy to see that (\ref{burban: qgener}), (\ref{burban: qfirsta}), (\ref{burban: qfirstb}), (\ref{burban: qforth})
at the corresponding values of the deformation parameters reduced to (\ref{burban: gener}),
(\ref{burban: firsta}), (\ref{burban: firstb}), (\ref{burban: gener2}), respectively.

In the following we consider the dependence of these algebras from each other.

\noindent {\it Case: (\ref{burban: qgener}) and (\ref{burban: qfirsta})  algebras.} The substitution  the values $a^+a$ and $aa^+$
from (\ref{burban: qgener}) into the left hand side of the first eq. (\ref{burban: qfirsta})  gives
\begin{equation}
aa^+-q^{l/\alpha} a^+a = [N + \frac{l}{\alpha\gamma}]_{pq}^{\alpha\gamma l} + [N]_{pq}^{\alpha\gamma l} = p^{-\alpha N}.
\end{equation}
The same take place for eq. (\ref{burban: qfirstb}). It means that with (\ref{burban: qgener})  implies (\ref{burban: qfirsta}) and
(\ref{burban: qfirstb}).
In order to prove the converse statement we find Casimir operator $C_1$ of the oscillator algebra (\ref{burban: qfirsta}).
It is easy to see that
\begin{equation}
[a, C_1] = 0,\quad [a^+, C_1] = 0,\quad [N, C_1] = 0,
\end{equation}
where
\begin{equation}\label{burban: inv1}
C_1 = p^{\alpha N}\Bigl(\frac{p^{-\alpha N} - q^{\gamma N}}{p^{-l/\gamma} - q^{l/\alpha}} - a^+a \Bigr).
\end{equation}
From here we have
\begin{equation} \label{burban: cas1} a^+a =  [N]_{pq}^{\alpha\gamma l} - p^{-\alpha N} C_1.
\end{equation}
The substitution of this expression  $a^+a $ into (\ref{burban: qfirsta}) gives
\begin{equation}\label{burban: cas2}
aa^+ = [N+\frac{l}{\alpha\gamma}]_{pq}^{\alpha\gamma l} - p^{-\alpha (N+\frac{l}{\alpha\gamma})} C_1,
\end{equation}
which coincides with (\ref{burban: qgener}) if $C _1 = 0.$

We will construct the representation of (\ref{burban: qfirsta}) in which $C_1$ is non equal to  zero.
To this we consider the representation of the relations (\ref{burban: qfirsta})
$$
a|n\rangle = q^{\gamma \nu_0/2}\Bigl(\frac{p^{-\alpha n} - q^{\gamma n}}{p^{-l/\gamma} - q^{l/\gamma}} \Bigr)^{1/2}|n - \frac{l}{\alpha\gamma}\rangle,
 a^+|n\rangle = q^{\gamma \nu_0/2}\Bigl(\frac{p^{-\alpha n -l/\gamma} - q^{\gamma n +l/\alpha}}{p^{-l/\gamma} - q^{l/\alpha}}\Bigr)^{1/2}|n +
\frac{l}{\alpha\gamma}\rangle,$$
\begin{equation}
N|n\rangle=(n+\nu_0)|n\rangle.
\end{equation}
In this representation \begin{equation} C_1|n\rangle = p^{\alpha \nu_0}[\nu_0]_{pq}^{\alpha\gamma l} |n\rangle\ne 0,
\end{equation}
if $\nu_0\ne 0.$
It means (\ref{burban: qfirsta}) does not imply (\ref{burban: qgener}).

\noindent {\it Case: (\ref{burban: qgener}) and (\ref{burban: qforth}) algebras}. Inserting in the left hand side of  (\ref{burban: qforth})  the values $a^+a$ and $aa^+$
from (\ref{burban: qgener}) we obtain identity. It means  that  (\ref{burban: qgener}) implies (\ref{burban: qfirsta}).
For the converse statement, take into account  the equation (\ref{burban: qforth}), we construct the Casimir operator $C_2$ of the algebra (\ref{burban: qforth})
\begin{equation}
[a, C_2] = 0,\quad [a^+, C_2] = 0,\quad [N, C_2] =  0,
\end{equation}
where
\begin{equation}\label{burban: casimir1}
 C_2 = (p^{\alpha} q^{-\gamma})^N ([N]_{pq}^{\alpha\gamma l} - a^+a).
\end{equation}
From this, we obtain
\begin{equation}\label{burban: cas3}
a^+a = [N]_{p,q}^{\alpha\gamma l} - (p^{-\alpha}q^{\gamma})^N C_2, \quad aa^+ = [N+\frac{l}{\alpha\gamma}]_{pq}^{\alpha\gamma l} - (p^{-\alpha}q^{\gamma})^{N+\frac{l}{\alpha\gamma}} C_2 .
\end{equation}
Non-equivalence (\ref{burban: qgener}) and (\ref{burban: qforth}) will be established  if we will prove $C_2\ne 0.$
For this we construct the representation of the algebra (\ref{burban: qforth}):
$$a^+|n\rangle = ([n+\frac{l}{\alpha\gamma} - \nu_0]_{pq}^{\alpha\gamma l}+[\nu_0]_{pq}^{\alpha\gamma l})^{1/2}|n+\frac{l}{\alpha\gamma}\rangle,\quad
 a |n\rangle = ([n -\nu_0]_{p,q}^{\alpha\gamma l}+[\nu_0]_{pq}^{\alpha\gamma l})^{1/2}|n - \frac{l}{\alpha\gamma}\rangle, $$
\begin{equation}
N|n\rangle =(n-\nu_0)|n\rangle,
\end{equation}
in which
$$ C_2|n\rangle = (p^{\alpha}q^{-\gamma})^n[\nu_0]_{pq}^{\alpha\gamma l}|n\rangle.$$ If  $\nu_0 \ne 0$ then $C_2\ne 0.$
It means that (\ref{burban: qforth})  in general does not imply (\ref{burban: qgener}).
\medskip

\noindent {\it Cases: (\ref{burban: qfirsta}) and (\ref{burban:
qforth}) algebras.} The substitution (\ref{burban: cas1}) and
(\ref{burban: cas2}) into the left hand side  of
 (\ref{burban: diff}) gives:
\begin{equation}
aa^+- (p^{-\alpha}q^{\gamma})^{\frac{l}{\alpha\gamma}} a^+a   =
[N+\frac{l}{\alpha\gamma}]_{pq}^{\alpha\gamma l} -
(p^{-\alpha}q^{\gamma})^{\frac{l}{\alpha\gamma}}
[N]_{p,q}^{\alpha\gamma l}-p^{-\alpha N+ \frac{l}{\alpha\gamma}}(1 -
q^{l/\alpha})C_1,
\end{equation}
which means that if $C_1\ne 0$, (\ref{burban: qfirsta}) does not imply (\ref{burban: qforth}).

Conversely, the substitution of (\ref{burban: cas3}) into the left hand side of the first eq.(\ref{burban: qfirsta}) gives
\begin{equation}
aa^+ - p^{-l/\gamma}a^+a = [N+\frac{l}{\alpha\gamma}]_{pq}^{\alpha\gamma l} - p^{-l/\alpha} [N]_{p,q}^{\alpha\gamma l} + C_2(p^{-\alpha}q^{\gamma})^N p^{-l/\gamma}(1- q^{l/\alpha})
\end{equation}
and into the left hand side of the first eq.(\ref{burban: qfirstb})
\begin{equation}
aa^+ - q^{l/\alpha} a^+a = [N+\frac{l}{\alpha\gamma}]_{pq}^{\alpha\gamma l} - q^{l/\alpha} [N]_{pq}^{\alpha\gamma l} + C_2(p^{-\alpha}q^{\gamma})^N q^{-l/\alpha}(1- p^{\gamma}),
\end{equation}
which means that (\ref{burban: qforth}) does not imply (\ref{burban: qfirsta}) if $C_2\ne 0.$
\medskip

\noindent
{\bf 3. Generalized $(p,q;\alpha,\gamma, l)$-deformed $U_{pq}^{\alpha\gamma l}(su(2))$ algebras}

\medskip
\noindent {\it Generalized  $(p,q;\alpha,\gamma, l)$-deformed
Jordan-Schwinger realization of $U_{pq}^{\alpha\gamma l} (su(2))$
algebra.} Consider two independent basis collections $(I,a,a^+ N_a)$
and $(I,b, b^+, N_b)$  of the algebra of the ordinary harmonic
oscillator. The Jordan-Schwinger transformations
\begin{equation}\label{burban: j-sh}
J_+= a^+b,\quad J_- = a b^+ , \quad J_0=\frac{1}{2}(N_a-N_b)
\end{equation}
give the realization of the $su(2)$ Lie algebra
\begin{equation}
[J_0, J_+] = J_+,\quad[J_0, J_-] = - J_-,\quad [J_+, J_-] = 2J_0.
\end{equation}
The two independent basis collections $(I, a, a^+ N_a)$ and $(I, b, b^+, N_b)$ of the generalized $(p,q; \alpha,\gamma, l)$-deformed oscillator  algebras define the generalized Jordan-Schwinger transformations
\begin{equation}\label{burban: shw}
J_+ = (p^{\alpha}q^{-\gamma})^{N_b/2}a^+b, J_- = b^+a(p^{\alpha}q^{-\gamma})^{N_b /2},
J_0 =\frac{1}{2}(N_a-N_b), {\tilde C} = \frac{1}{2}(N_a + N_b).
\end{equation}
In the case of collections of the generalized $(p,q; \alpha,\gamma, l)$-deformed Chakrabarti-Jagannathan type oscillator algebra (\ref{burban: qfirsta}) (or (\ref{burban: qfirstb})),
we obtain the $U_{pq}^{\alpha \gamma l}\Bigl(su(2)\Bigr)$ generaized algebra with commutation relations
\begin{equation}
[J_0, J_{\pm}] = \pm J_{\pm},\quad J_+J_-
-(p^{\alpha}q^{-\gamma})^{\frac{l}{\alpha\gamma}}J_- J_+ = \bigl(1-
C_1(p^{-l/\gamma} - q^{l/\alpha}))\bigr [2 J_0]_{pq}^{\alpha\gamma
l},
\end{equation}
and, in the case collections  of the $(p,q; \alpha,\gamma, l)$-deformed Hong Yan type oscillator algebra (\ref{burban: qforth}), the $U_{pq}^{\alpha\gamma l}\Bigl(su(2)\Bigr)$ generalized  deformed algebra with the commutation relations
$$ [J_0, J_{\pm}] = \pm J_{\pm}, \quad J_+J_- -(p^{\alpha}q^{-\gamma})^{\frac{l}{\alpha\gamma}}J_- J_+ = [2J_0]_{pq}^{\alpha\gamma l} +$$
\begin{equation}
 (p^{\alpha} q^{-\gamma})^{{\tilde C} - J_0}C_2\bigl( [{\tilde C}- J_0 +\frac{l}{\alpha\gamma}] -(p^{-\alpha}q^{\gamma})^{2J_0}[{\tilde C} - J_0]-(p^{-\alpha}q^{\gamma})^{2{\tilde C}}
 [{\tilde C}+J_0+\frac{l}{\alpha\gamma}]+(p^{-\alpha}q^{\gamma})^{\frac{l}{\alpha\gamma}}[{\tilde C}+J_0].
\end{equation}

\medskip
\noindent {\it Generalized  $(p,q;\alpha,\gamma, l)$-deformed
Holstein-Primakoff realization of $U_{pq}^{\alpha\gamma l} (su(2))$
algebra.} The algebra $su(2)$ can be realized by  the one basis
collection  $(a, a^+, N_a)$ of the operators of the ordinary
harmonic oscillator. It is defined by the Holstein-Primakoff
transformations
\begin{equation}
  J_+ = a^+(2 j - N)^{1/2},\quad J_- = (2j - N)^{1/2}a,\quad J_0 = N - j,
\end{equation} where $j$ is a $c$-number.

The $q$-deformed analog Holstein-Primakoff  of the algebra $su(2)$ has been studied in \cite{CEK}.
The generalized Holstein-Primakoff realization of the algebra $su(2)$ defined by the collection of the $ (p,q; \alpha,\gamma, l)$-deformed oscillator is given by
\begin{equation}\label{burban: hp}
J_+ = (p^{\alpha}q^{-\gamma})^{N/2}a^+\sqrt{[2j - N]_{pq}^{\alpha\gamma l}}, J_- =\sqrt{[2j - N]_{pq}^{\alpha\gamma l}} a (p^{\alpha}q^{-\gamma})^{N /2},
J_0=N - j,
\end{equation}
where $j$ is some $c$-number.
For the collection of the generalized Chakrabarti-Jagannathan oscillator algebra (\ref{burban: qfirsta})
we obtain the  $U_{pq}^{\alpha \gamma l}\Bigl(su(2)\Bigr)$ algebra with the relations
\begin{equation}
[J_0, J_{\pm}] = {\pm} J_{\pm},\quad J_+J_-
-(p^{\alpha}q^{-\gamma})^{\frac{l}{\alpha\gamma}}J_- J_+ =  [- 2
J_0] + Cq^{-2\gamma J_0}.
\end{equation}

\medskip

{\bf 4. The $ (p,q; \alpha,\gamma, l)$-deformation of two-dimensional conformal field theory}

\medskip

\noindent {\it Some aspects of  $ (p,q; \alpha,\gamma, l)$-deformed CFT in view of standard CFT.} We will survey the basic features of the standard
CFT \cite{BPZ}.

The transformations  $x^a\mapsto x'^a $  for which
\begin{equation} g'_{\rho\sigma}(x')\frac{\partial x'^{\rho}\partial x'^{\sigma}}{\partial x^{\mu}\partial x^{\nu}} = \Lambda(x) g_{\mu \nu}(x), \end{equation}
where $g_{\mu\nu}(x)$ is metric tensor and $\Lambda (x)$ positive function  (scale factor), form the conformal group.
 Among all fields in the CFT there are  distinguished ones (primary fields of a conformal weights $(h, {\bar h}),$ ) i.e., the fields, which  behave as tensors
\begin{equation}
\phi(z',{\bar z}')=\phi(z,{\bar z})(\frac{dz'}{dz})^{-h}(\frac{d{\bar z}'}{d{\bar z}})^{-{\bar h}}\end{equation}
under the conformal transformations
$z\to z' = f(z),   {\bar z}\to {\bar z}' = {\bar f}({\bar z}).$
The infinitesimal form of these transformations $f(z) = z+\varepsilon(z)\quad {\bar f}(\bar z) = {\bar z} +{\bar \varepsilon}({\bar z})$
has the form
\begin{equation} \label{burban: trans1}
\phi'(z,{\bar z})\mapsto\phi(z,{\bar z})+\Delta_{\varepsilon}\phi(z, {\bar z}),
\end{equation}
where
\begin{equation}
\Delta_{\varepsilon}\phi(z, {\bar z} = \delta_{\varepsilon}\phi(z,\bar{z})+{\bar \delta}_{\bar{\varepsilon}}\phi(z,\bar{z}),
\end{equation}
and
$$
\delta_{\varepsilon}\phi(z,\bar{z}) = \varepsilon(z)^{1-h}\partial_{ z}\Bigl( \varepsilon(z)^h\phi(z,\bar z)\Bigr),
$$
\begin{equation}\label{burban: trans2}
{\bar \delta}_{\bar{\varepsilon}}\phi(z,\bar{z})= {\bar \varepsilon}(\bar z)^{1-{\bar h}}\partial_{\bar z}\Bigl( {\bar\varepsilon}(\bar z)^{\bar h}\phi(z,\bar z)\Bigr)
\end{equation}

In particular, when $\varepsilon(z) = \varepsilon_n z^{n+1}$ end ${\bar \varepsilon}({\bar z}) = {\bar \varepsilon}_n {\bar z}^{n+1},$ where
${\varepsilon}_n$ and ${\bar \varepsilon}_n $ are small constants we have
$$
\phi'(z,{\bar z}) = \phi(z,{\bar z})+\Delta_n\phi(z,{\bar z}),
$$ where
 \begin{equation}\Delta_n\phi(z,{\bar z}) =\varepsilon_n\delta_n \phi(z,{\bar z}) +  {\bar \varepsilon}_n{\bar \delta}_n \phi(z,{\bar z})\end{equation}
and
$$
\delta_n\phi(z,{\bar z}) = z^n (z\partial+h(n+1))\phi(z,{\bar z}),$$
\begin{equation}\label{burban: ctransfor}
{\bar \delta}_n\phi(z,{\bar z}) ={\bar z}^n ({\bar z}{\bar \partial}+h(n+1))\phi(z,{\bar z}).
\end{equation}

The different deformation schemes of the oscillator algebras have been transfer, in parallel, to  the construction of two-dimensional conformal field theories \cite{Mat,  Ber, OhS, Chu}.
We study some properties of  primary fields of the two-dimensional  deformed conformal field theory with the employment of the $(p,q;\alpha,\gamma, l)$-deformation
scheme.

We leave the same notation $\phi(z,\bar z),$ as in classical case, for the primary field of the conformal weights $(h,\bar h)$ of the $(p,q;\alpha,\gamma, l)$-deformed conformal field theory.
Its generalized infinitesimal $ (p,q; \alpha,\gamma, l)$-transformations are defined by replacement in (\ref{burban: trans2}) the $\partial_z$ by its $(p,q; \alpha,\gamma, l)$-deformed
counterpart
\begin{equation}
\delta_{\varepsilon, pq}^{\alpha\gamma l}\phi (z) = {\varepsilon}(z)^{1-h}D^{\alpha\gamma l}_{pq}(\varepsilon(z))^h \phi (z)),\end{equation}
where $$ D^{\alpha\gamma l}_{pq}\phi(z)= \frac{\phi(p^{-\alpha} z) - \phi(q^{\gamma} z)}{z(p^{-l/\gamma} - q^{l/\alpha})}
= \frac{1}{z}.\frac{p^{-\alpha z\partial}-q^{\gamma z\partial}}{p{-l/\alpha}-q^{l/\gamma}}\phi(z)=\frac{1}{z}[z\partial]_{pq}^{\alpha\gamma l}\phi(z).$$
Henceforth we will restrict our consideration by the holomorphic terms.

The $ (p,q; \alpha,\gamma,  l)$-deformed counterpart of the transformation (\ref{burban: ctransfor})  for primary field $\phi (z)$ with the conformal weight  $h$ is
\begin{equation}\label{burban: quanta}\delta_{n,pq}^{\alpha\gamma l}\phi(z) =  z^n [z\partial+h(n+1)]^{\alpha\gamma l}_{pq}\phi(z).
\end{equation}
In quantum field theory a variation of the field $\phi (z)$
is implemented by the commutator $$ \delta A = [Q, A],$$
where $$Q=\int dx^1 j_0, \qquad x^0 = const $$ and $j_0$ is conserved current.
In our case the variation conformal field $\phi(z)$ is given by the "equal -time" commutator
\begin{equation}\label{burban: prod}
\delta_{n,pq}^{\alpha\gamma l}\phi(z) =
[\oint_{C_0}\frac{dz}{2\pi i}z^n T(z),\phi(w)] =
 \frac{1}{2\pi i} \oint _{C_P} dz  z^{n+1}R\Bigl( T(z) \phi(w) \Bigr)_{p,q}^{\alpha\gamma l},
\end{equation}
where $T(z)$ is the holomorphic component of the energy momentum tensor and $C_0$ and $C_P$ is a contous encircling the origin and  all poles in the OPE of $\Bigl(T(z)\phi(w\Bigr)_{pq} ^{\alpha\gamma l}$ respectively.
 The notation of the time ordering on the $z$ plane is replaced by that of radial ordering
$$
R(A(z)B(w))=
\cases{
A(z)B(w), &if $|z|>|w|,$\cr
B(w)A(z), &if $|z|<|w|$.\cr}.
$$
It is easy to see that
\begin{equation}
\Bigl( T(z) \phi(w) \Bigr)_{pq}^{\alpha\gamma l} =
\frac{1}{w (p^{-l/\gamma} - q^{l/\alpha})} \{ \frac{\phi (w p^{-\alpha}) } {z - w p^{-\alpha h }}  -
\frac{\phi (wq^{\gamma})}{z- w q^{\gamma  h}} \}
\end{equation}
leads to the correct of the variation (\ref{burban: quanta}) in
$\phi(z)$ by evaluating integral in eq.(\ref{burban:
prod}) with $C_P$ taken as contour encircling the points ${w
p^{-\alpha h} ,w q^{\gamma  h}}.$

Introducing the modes
\begin{equation}
L_n =\oint_{C_0}\frac{dz}{2\pi i}z^{n+1}T(z),\quad n\in {\mathbb Z}
\end{equation} of a holomorphic component of the energy momentum tensor $T(z)$
and the modes
\begin{equation}
\phi_n=\oint_{C_0}\frac{dw}{2\pi i}w^{n+h-1}\phi(w)
\end{equation}
 of a primary field $\phi(w)$ of the conformal weight $h,$
we can  evaluate the bracket
$$[L_n,\phi_m ] = (L_n\phi_m)^{\alpha\gamma l}_{pq}- (\phi_m L_n)^{\alpha\gamma l}_{pq}. $$
Defining the product of two field operators by
$$\Bigl(A(z)B(w)\Bigr)_{pq}^{\alpha\gamma l} = A(zq^{\gamma})B(wp^{-\alpha}) $$
we obtain
$$ (L_n\phi_m)^{\alpha\gamma l}_{pq} =
\oint _{C_1}\frac{ dz}{2\pi i}\oint _{C_2}\frac{ dw}{2\pi i}
z^{n+1}w^{m+h-1}(T(z) \phi(w))^{\alpha\gamma l}_{pq}=$$$$
\oint _{C_1}\frac{ dz}{2\pi i}\oint _{C_2}\frac{ dw}{2\pi i}
z^{n+1}w^{m+h-1}(T(z q^{\gamma}) \phi(wp^{-\alpha}))^{\alpha\gamma l}_{pq}=$$
$$ \oint _{C_1}\frac{ dz}{2\pi i}\oint _{C_2}\frac{ dw}{2\pi i}
z^{n+1}w^{m+h-1}\sum_k L_k z^{-k-2}q^{-\gamma(k+2)}\sum_l \phi_l (w
p^{\alpha(l+h)}) w^{-l - h}= $$
 \begin{equation}
 p^{\alpha(m+h)}q^{-\gamma(n+2)}L_n\phi_m,
  \end{equation}
 where $C_1$ and $C_2$ are contours about the origin such that $C_1\subset C_2.$
 Analogously,
\begin{equation} (\phi_m L_n)_{pq}^{\alpha\gamma l}  = p^{\alpha(n+2)}q^{-\gamma(m+h)}\phi_m L_n.\end{equation}
Hence, this bracket has been written as
\begin{equation}
[L_n,\phi_m] := p^{\alpha(m+h))}q^{-\gamma(n+2)} L_n\phi_m- p^{\alpha(n+2)}q^{-\gamma(m+h)}\phi_m L_n.
\end{equation}
To evaluate this bracket we use the $(p,q; \alpha\gamma l)$-deformed OPE
$$
[L_n,\phi_m] = \oint _{C_0}\frac{ dz}{2\pi i}\oint _{C_P}\frac{ dw}{2\pi i}
 z^{n+1}w^{m+h-1}R(T(z) \phi(w,{\bar w}))_{pq}^{\alpha\gamma l} =
$$
$$
\frac{1}{(p^{-l/\gamma}-q^{l/\alpha})}
\oint _{C_0}\frac{dz}{2\pi i }\oint_{C_P} \frac{ dw}{2\pi i}
z^{n+1}w^{m+h-1}\{ \frac{\phi (w p^{-\alpha})}{z- w p^{-\alpha  h}}
 - \frac{\phi (wq^{\gamma})}{z- w q^{\gamma h}} \}  =
$$
$$
\frac{1}{(p^{-l/\gamma} - q^{l/\alpha})} \oint_{C_P} \frac{dw}{2\pi i} w^{n+1} w^{m+h-2}\times $$
$$
\{ p^{-{\alpha h}(n+1)} \sum_k \phi_k p^{\alpha(k+h)}\omega^{-k-h} - q^{\gamma h(n+1)}\sum_k \phi_k q^{- \gamma (k+h)}\omega^{-k -h} \}  =
$$
$$
\frac{1}{(p^{-l/\gamma} - q^{l/\alpha)}}\{p^{-\alpha [(h-1)n - m]} - q^{\gamma[(h-1)n -m]}\}\phi_{m+n} =
$$
$$
 [(h-1)n -m]^{\alpha\gamma l}_{pq}\phi_{n+m}.
 $$
 Hence "quommutator" $L_n$ and $\phi_m$ is given
\begin{equation}
 p^{\alpha(m+h))}q^{-\gamma(n+2)} L_n\phi_m-
 p^{\alpha(n+2)}q^{-\gamma(m+h)}\phi_m L_n = [(h-1)n -m]^{\alpha\gamma l}_{pq}\phi_{m+n}.
\end{equation}
If in this equation we let $\phi_m = L_m,$ $h=2$ we obtain the centerless $(p,q; \alpha,\gamma, l)$ - deformed Virasoro algebra
\begin{equation}
p^{\alpha(m+2))}q^{-\gamma(n+2)} L_n L_m-
p^{\alpha(n+2)}q^{-\gamma(m+2)}L_m L_n = [n -m]^{\alpha\gamma}_{pq}L_{m+n}.
\end{equation}
which at the $\alpha=\gamma= l $ and $p=q$ coincides with the algebra obtained in \cite{CZ}.

The space of states of a two-dimensional conformal field theory carries the representation of the Virasoro algebra
\begin{equation}
[{\bar L}_m,{\bar L}_n]=(m-n){\bar L}_{m+n}+\frac{c}{12}n (n+1)\delta_{m+n, 0}\quad n = 1,2,\ldots
\end{equation}
This algebra contains of the closed $sl(2,{\mathbb C})$ subalgebra
$$[{\bar L}_0, {\bar L}_{-1}] = {\bar L}_{-1},\quad {\bar L}_0,{\bar L}_{+1}] =
-{\bar L}_{+1},\quad [{\bar L}_{-1},{\bar L}_{+1}] = - 2 {\bar
L}_0.$$ One can introduce the $*$-structure in this algebra $$ {\bar
L}_0^* ={\bar L}_0, \quad {\bar L}_{+1}^* = - {\bar L}_{-1},\quad
{\bar L}_{-1}^* = - {\bar L}_{+1}  $$ which turn this algebra into
the $su(1,1)$ Lie algebra which after the renaming ${\bar L}_0 = -
K_0,\quad {\bar L}_{-1} =  K_{+1},\quad {\bar L}_{+1}= K_{-1}$
 satisfy by the commutation relations
$$[K_{-1},K_{+1}] = 2 K_0,\quad [K_0,K_{-1}]= - K_{-1},\quad [K_0,K_{+1}] = K_{+1}.$$
The $(p,q)$-quantum deformation and the co-algebra structure of this $U_{pq}$(su(1,1 algebra  has been constructed in \cite{Ch-J}.

\noindent {\it Representation of $U_{pq}^{ \alpha\gamma l}\Bigl(su(1,1)\Bigr)$-deformed  algebra.}
The generalize $U_{pq}^{ \alpha\gamma l}\Bigl(su(1,1)\Bigr)$-deformed  algebra is defined by the commutation relations
$$ K_- K_+- (p^{\alpha}q^{-\gamma})^{\frac{l}{\alpha\gamma}}K_+K_- =
= [2K_0]_{pq}^{\alpha\gamma l},$$
\begin{equation} [K_0, K_+] = K_+,\quad  [K_0, K_-] = - K_-,
\end{equation}
where $ [2K_0]_{pq}^{\alpha\gamma l} = \Bigl(p^{-2\alpha K_0 } -
q^{2 \gamma K  _0}\Bigr)/(p^{-l/\gamma} - q^{l/\alpha}).$ Moreover,
it is convenient  to define the generators
$$ M = p^{-\alpha K_0}, N = q^{\gamma K_0}$$ .

There exists the following representation of this algebra on the conformal fields $\phi(z)$ with conformal weight $h$
$$ K_{-1}\phi(z) = \frac{1}{z} \frac{\phi(p^{-\alpha}z) - \phi (q^{\gamma }z)}{ p^{-l/\gamma} - q^{l/\alpha} },\quad
K_{+1} \phi(z) = z \frac{p^{-2\alpha h}\phi(p^{-\alpha} z) - q^{2\gamma h}\phi(q^{\gamma} z)}{ p^{-l/\gamma} - q^{l/\alpha}},$$
\begin{equation}
M \phi(z) = p^{-\alpha h}\phi(p^{-\alpha}z),\quad N = q^{\gamma
h}\phi(q^{\gamma}z).\end{equation}
The co-algebra structure of this algebra is defined on the generators of the algebra by
the expressions
\begin{equation}
\Delta(K_{\pm}) = M\otimes K_{\pm}+K_{\pm}\otimes N,\quad \Delta(M) = M\otimes M,\quad \Delta(N) = N\otimes N.
\end{equation}
\noindent {\it Two-point correlation function of $ (p,q; \alpha,\gamma, l)$-deformed CFT.} By analogy with \cite{Mat, Ber} the $ (p,q; \alpha,\gamma,l)$-deformed Ward-Takahashi identities for $n$-point correlation function  of the primary fields $\phi_1(z_1), \phi_2(z_2),\ldots,\phi_n(z_n)$ can be written
\begin{equation}
 \overbrace{(\Delta\otimes id \ldots\otimes id)}^{N-1}\overbrace{(\Delta\otimes id \ldots\otimes id)}^{N-2}\ldots(\Delta\otimes id)
\Delta(K_{\pm})\langle(\phi_1(z_1)\ldots\phi_N(z_N) \rangle_{pq} ^{\alpha\gamma l}= 0.
\end{equation}
For the two-point correlation unction we have the equations
$$\Delta (K_{-1})\langle\phi_1(z_1)\phi_2 (z_2)  \rangle_{pq}^{\alpha\gamma l } = 0 ,\quad
\Delta(K_{+1})\langle\phi_1(z_1)\phi_2 (z_2)  \rangle_{pq}^{\alpha\gamma l } = 0, $$
or in expanded form
$$p^{-\alpha h_1}\frac{1}{z_2}[ \langle \phi_1(p^{-\alpha}z_1)\phi_2(p^{-\alpha }z_2)\rangle_{pq}^{\alpha\gamma l } - \langle\phi_1(p^{-\alpha}z_1)\phi_2 (q^{\gamma} z_2)\rangle_{pq}^{\alpha\gamma l }] + $$
\begin{equation}\label{burban: corr1}
q^{\gamma h_2}\frac{1}{z_1}[
\langle\phi_1(p^{-\alpha}z_1)\phi_2(q^{\gamma}
z_2)\rangle_{p,q}^{\alpha\gamma l }- \langle\phi_1(q^{\gamma}
z_1)\phi_2 (q^{\gamma}z_2)\rangle_{pq}^{\alpha\gamma l }] = 0,
\end{equation}
$$
 p^{-\alpha h_1}z_2[ p^{-2\alpha
h_2}\langle\phi_1(p^{-\alpha}z_1)\phi_2(p^{-\alpha}z_2)
\rangle_{pq}^{\alpha\gamma l } - q^{-2\gamma h_2}
\langle\phi_1(p^{-\alpha}z_1)\phi_2(q^{\gamma}z_2)\rangle_{pq}^{\alpha\gamma
l }] +
 $$
\begin{equation}\label{burban: corr2}
 q^{\gamma h_2} z_1[ p^{-2\alpha h_1}\langle\phi_1(p^{-\alpha} z_1)
 \phi_2(q^{\gamma} z_2) \rangle_{pq}^{\alpha\gamma l }-
 q^{2\gamma h_2}\langle\phi_1(q^{\gamma}z_1)\phi_2(q^{\gamma} z_2)
 \rangle_{pq}^{\alpha\gamma l }] = 0.
\end{equation}
The eq. (\ref{burban: corr1}) can be represented in the form
$$
(\frac{ q^{\gamma h_2}}{z_1} - \frac{p^{-\alpha h_1}}{z_2})\langle\phi_1(p^{-\alpha} z_1)\phi_2 (q^{\gamma}z_2)\rangle_{pq}^{\alpha\gamma l }+$$
$$
\frac{p^{-\alpha h_1}}{z_2} \langle \phi_1(p^{-\alpha}z_1)\phi_2(p^{-\alpha }z_2)\rangle_{pq}^{\alpha\gamma l }-\frac{q^{\gamma h_2}}{z_1}\langle\phi_1(q^{\gamma} z_1)\phi_2 (q^{\gamma}z_2)\rangle_{pq}^{\alpha\gamma l } = 0.
$$
And if $h_1 = h_2$ we have
$$
(1 -\frac{(p^{\alpha}q^{\gamma})^{-h}}{z_2/z_1})\langle\phi_1(p^{-\alpha} z_1)\phi_2 (q^{\gamma}z_2)\rangle_{pq}^{\alpha\gamma l }+$$
\begin{equation}\label{burban: corr3}
\frac{(p^{\alpha}q^{\gamma})^{-h}}{z_2/z_1}\langle \phi_1(p^{-\alpha}z_1)\phi_2(p^{-\alpha }z_2)\rangle_{pq}^{\alpha\gamma l }-
\langle\phi_1(q^{\gamma} z_1)\phi_2 (q^{\gamma}z_2)\rangle_{pq}^{\alpha\gamma l } = 0.
\end{equation}
We will find the solution of this equation using the following anzatz:
\begin{equation}\label{burban: solut}
\langle\phi_1( z_1)\phi_2 (z_2)\rangle_{pq}^{\alpha\gamma l } =
z_1^{\omega}h_a(z)  = z_1^{\omega}\frac{(az; r)_{\infty}}{(z;
r)_{\infty}},
\end{equation}
where $ a =(p^{\alpha}q^{\gamma} )^{2h}, z=  (p^{\alpha}q^{\gamma} )^{-h}\frac{z_2}{z_1},r =p^{\alpha}q^{\gamma},\omega \in{\mathbb R}.$

Then
\begin{equation}\label{burban: corr4}
\langle\phi_1(p^{-\alpha} z_1)\phi_2 (q^{\gamma}z_2)\rangle_{pq}^{\alpha\gamma l } = p^{-\alpha \omega}\frac{1-(p^{\alpha}q^{\gamma})^{-h}z_2/z_1}{1-(p^{\alpha}q^{\gamma})^{-h}z_2/z_1}\langle\phi_1(z_1)\phi{z_2}(z_2)\rangle_{pq}^{\alpha\gamma l}.
\end{equation}
We have used the relation $$h_a(z)= \frac{1-az}{1-z}h_a(qz) $$ (see (1.3.11) in \cite{GR})
to obtain (\ref{burban: corr4}).
In addition,
\begin{equation}\label{burban: corr5}
\langle \phi_1(p^{-\alpha}z_1)\phi_2(p^{-\alpha }z_2)\rangle_{pq}^{\alpha\gamma l }=p^{-\alpha\omega}\langle\phi_1(z_1)\phi{z_2}(z_2)\rangle_{pq}^{\alpha\gamma l }
\end{equation}
and
\begin{equation}\label{burban: corr6}
\langle\phi_1(q^{\gamma} z_1)\phi_2 (q^{\gamma}z_2)\rangle_{pq}^{\alpha\gamma l } =q^{\alpha \omega}\langle\phi_1(z_1)\phi{z_2}(z_2)\rangle_{pq}^{\alpha\gamma l }.
\end{equation}
The direct substitution of (\ref{burban: corr4}), (\ref{burban: corr5}),(\ref{burban: corr6}) into (\ref{burban: corr3})  gives $\omega= - 2 h$ and define the two-point correlation function  (\ref{burban: solut}). It also satisfy the eq. (\ref{burban: corr2}).

\medskip

\noindent {\bf 5. Summary and conclusions}

\medskip

In this letter we have presented the generalized deformation of the oscillator algebra obtained by the using the generalization of structure function
of the deformations  well- known  Arik-Coon, Biedenharn-Macfarlane, Jagannathan-Chakrabarti deformer algebras                                                                                                                 and as well as exchanging the defining relations of these algebras. The each of these as well as non-deformed algebras are obtained as particular case
of the generalized algebra at special choice parameters of structure function. The  unification conserve basic properties of these deformed algebras.
We have find the Biedenharn-Macfarlane and Holstein-Primakoff realizations $U_{pq}^{\alpha\gamma l}\Bigl(su(2)\Bigr)$ algebra by the generalized creation and
annihilation operators.

This generalized $(p,q;\alpha,\gamma, l)$-deformation of
 the oscillator algebra is "the attempt to introduce  some order  in the rich and varied choice of deformed commutation relations"
\cite{Q1} and also the hope that it will find applications in the solutions of the concrete physical problems \cite{Bal}.
The  same concerns of the two-dimensional generalized deformed conformal field theories which special cases have been studied in \cite{ Mat, Ber, OhS, Chu}.
We have presented  the unified  $ (p,q; \alpha,\gamma,  l)$-deformed version of the two-dimension conformal field theory and have calculated  two-point correlation function
of its primary fields.

\medskip
\noindent
{\bf Acknowledgments}
\medskip

This research was partially supported by the Special Program of the Division of Physics and Astronomy of NAS of Ukraine.
I would like to thank the referees for the granted useful information and pointed out the mistakes and the misprints in the original manuscript.

\end{document}